**Susceptibility functions for slow relaxation processes in supercooled liquids and the search for universal relaxation patterns**

Th. Blochowicz, Ch. Tschirwitz, St. Benkhof, E. A. Rössler

Experimentalphysik II, Universität Bayreuth, D 95440 Bayreuth

Abstract

In order to describe the slow response of a glass former we discuss some distribution of correlation times, e.g., the generalized gamma distribution (GG) and an extension thereof (GGE), the latter allowing to reproduce a simple peak susceptibility such as of Cole-Davidson type as well as a susceptibility exhibiting an additional high frequency power law contribution (excess wing). Applying the GGE distribution to the dielectric spectra of glass formers exhibiting no β-process peak (glycerol, propylene carbonate and picoline) we are able to reproduce the salient features of the slow response ($10^{-6}$ Hz – $10^{9}$ Hz). A line shape analysis is carried out either in the time or frequency domain and in both cases an excess wing can be identified. The latter evolves in a universal way while cooling and shows up for correlation times $\tau_\alpha > 10^{-8}$ s. It appears that its first emergence marks the break down of the high temperature scenario of mode coupling theory. – In order to describe a glass former exhibiting a β-process peak we have introduced a distribution function which is compatible with assuming a thermally activated process in contrast to some commonly used fit functions. Together with the GGE distribution this function allows in the frame of the Williams-Watts approach to completely interpolate the spectra, e.g. of fluoro aniline ($10^{-6}$ Hz – $10^{9}$ Hz). The parameters obtained indicate an emergence of both the excess wing and the β–process again at $\tau_\alpha > 10^{-9}$ s.

## 1. Introduction

Upon supercooling a simple liquid the dynamic susceptibility evolves in a characteristic way. Above the melting point $T_m$ the main relaxation process, i.e., the structural relaxation or α–process occurs on the ps time scale and this relaxation slows down significantly when supercooling. Finally, close to the glass transition temperature $T_g$ structural relaxation times of some hundred seconds are reached which lead to structural arrest at somewhat lower temperatures. Here, the system does not relax any longer on the time scale of the experiment, and a glass is formed. Between the α–process time scale and the microscopic band (and boson peak) in the THz range a broad frequency window opens while cooling which is filled by the emergence of additional relaxation processes which are called secondary relaxation processes. Fig. 1 sketches the situation. It is one challenge of studying the glass phenomenon to provide a coherent description of the evolution of the dynamic susceptibility, another to understand the physical nature of the processes involved.

*Note: The appendix of the present paper has been omitted due to space limitation. Please contact authors for further information.*

Since the advent of mode coupling theory (MCT) significant experimental progress has been made in monitoring the evolution of the molecular dynamics [1]. For example, the dielectric response of some glass formers has now been measured over 19 decades in frequency ($10^{-6}$ Hz – $10^{13}$ Hz) [2]. Quasi-elastic light scattering is able to monitor the fast dynamics with high precision ($10^{8}$ Hz – $10^{13}$ Hz) [3-8] and neutron scattering experiments have compiled the corresponding q-dependence [9,10]. Together with molecular dynamics simulations [11] the application of the aforementioned techniques among others have demonstrated that MCT provides a rather convincing picture of the evolution of the susceptibility at the onset of the glass transition, i.e., at say $T > 1.2\ T_g$. In addition to the α–process a fast relaxation process ("fast dynamics" in Fig. 1) was identified. This feature is observed for organic as well as inorganic systems in a remarkable similar manner [12] and within MCT is attributed to some in-cage motion whereas the α–process describes the reorganization of the cage which enables flow. A MCT analysis allows for determining a critical temperature $T_c$ above which the high temperature scenario of the (idealized) theory accounts for the salient feature of the susceptibility [1]. Although not expected recent experiments show that this scenario is found even up to $T_m$ [7,8].

Below $T_c$ the experimental situation is less clear. The reason for this is mainly due to the emergence of slow secondary relaxation processes with small relaxation strengths. Two processes are discussed (cf. Fig. 1). First, an excess high frequency contribution to the α–relaxation peak appears while cooling in many glass formers, a feature well known since the work of Davidson and Cole [13]. For its evolution, Nagel and coworkers proposed a scaling procedure including both the α–process peak and the high frequency wing [14]. Though the scaling procedure suffers from formal drawbacks [15,16] the approach stresses the fact that the susceptibility exhibits high similarity among the different systems. We recently suggested to call such glass formers that only show an excess wing together with a α–process peak type A glass formers [17]. Below $T_g$ the excess wing contribution degenerates to a rather flat background which may be described by a nearly constant loss behavior in fair approximation [17,18]. Several proposals were offered to interpolate the full slow response of such type A glass formers but none is broadly accepted [17,19,20].

Second, in addition to the excess wing the Goldstein-Johari (JG) β–process is identified in many glass formers (cf. Fig. 1): Here a distinct second relaxation peak shows up in the Hz – kHz regime which persist in the glass ($T < T_g$). We classified these systems as type B glass formers [17]. Since the JG process is also found in glasses formed by rigid molecules it has to be regarded as an intrinsic relaxation process as already claimed by Johari and Goldstein [21]. Though the relaxation process exhibits a number of common features its relaxation strength strongly varies among the different systems. For example, toluene exhibits a strong secondary relaxation peak [17], whereas fluoro aniline shows a small one [17], and 2-picoline, a systems with the same molecular weight as toluene, displays no such peak [8]. Recently a discussion started addressing the question whether the wing contribution may be regarded as a secondary relaxation process, in particular, as a JG process [22-24].

Up to now it is neither understood how the emergence of the slow secondary relaxation processes is related to the anticipated change of the dynamics around the critical temperature $T_c$ nor do there exist theoretical predictions concerning these processes. Also, not much is known about the molecular motion taking place. Recent solvation [25] and NMR experiments [26,51] point out that the JG β–process involves small angle reorientation of essentially all molecules. However, this statement is recently challenged again by Johari [52]. Concerning the wing contribution its nature is even less understood. NMR experiments seem to indicate that it again can be attributed to some highly restricted motion [27,28].

For the time being, we are left with task to find a phenomenological description of the evolution of the dynamic susceptibility at $T < T_c$, and one may hope that the discussion of the parameters provides some insight on the change of the susceptibility upon cooling. We note that since no full description of the dynamic susceptibility has been offered so far any statement concerning, e.g., the validity of the frequency temperature superposition principle is not well founded since the fit results depend on the fitting interval. It is the purpose of this contribution to propose a complete description for the susceptibility spectra of the slow dynamics including α– and all secondary relaxation processes.

In order to perform a satisfying description of the response of a glass former well adapted interpolation formulae are needed. Often the main relaxation in supercooled liquids is described by applying analytic functions either in the time domain (e.g. Kohlrausch-Williams-Watts) or in frequency domain (e.g. Havriliak-Negami) [29]. A more general approach rather starts with introducing a distribution of correlation times $G(\ln\tau)$. Many distributions and susceptibility functions have been introduced [29], but most of them do not contain sufficient freedom to take proper account of the various spectral features such as the excess wing. Here we want to discuss distributions of correlation times which i.) allow to interpolate the α–process in cases when that process becomes very broad as it may be the case in polymers or in binary glass formers (section 2.1); ii.) allow to fully reproduce the slow dynamics in type A systems, i.e., allow to interpolate both α–peak and its high frequency wing (section 2.2); iii.) allow to reproduce the susceptibility of secondary relaxation processes, e.g., of JG type, in particular, if the spectra are asymmetric or may degenerate to some kind of excess wing (section 2.3). The distribution functions have to fulfill certain constraints for being physically reasonable. In the case of the α–process distribution we require that a mean correlation time exists, and in the case of the JG process we request that the distribution is compatible with, but not restricted to thermally activated dynamics. For example, the Havriliak-Negami function satisfies neither of these criteria, although it is often applied. For interpolating the full susceptibility including α–process, high frequency wing and JG process contribution we make use of the Williams-Watts approach [30].

Our starting point is a generalized gamma (GG) distribution which will be extended (GGE) to include the wing phenomenon. Some of the properties and applications of the GG distribution have already been discussed by Nicolai et al. [31] and of its extension by Kudlik et al. [17]. The proposed distribution GGE allows to interpolate an α–relaxation spectrum with and without high frequency wing. Furthermore, we present a distribution function ($G_\beta$) which may be applied to fit the β-process properly. In section 3 we proceed to some applications. We will present a complete interpolation of the dielectric loss at $T > T_g$ of a series of type A as well as of type B glass formers in the frequency range $10^{-6}$ Hz – $10^{9}$ Hz. In addition to previously reported data [17] we have included new data measured with a time domain spectrometer [32]. We will readdress the question up to what extent a universal evolution of the distribution parameters is found when a liquid is supercooled. Here, it appears favorable to discuss the parameters rather as a function of the time constant $\tau_\alpha$, than plotting them as a function of temperature, as it is usually done. In this way one can analyze the molecular slowing down independent of the degree of fragility. We will give new foundation for the idea [16] that the excess wing appears only above a certain time constant $\tau_\alpha$, the latter quantity being rather universal, which marks the sole change in the susceptibility spectra of type A glass formers when transforming from the liquid to a glass.

## 2 Distribution of correlation times

### 2.1 Distribution for the α–relaxation peak

The starting point of our discussion is to introduce a distribution G(lnτ) which is able to describe the various types of α–relaxation peaks. We propose to apply a generalized gamma distribution (GG) [31]

$$G_{GG}(\ln\tau) = N_{GG}(\alpha,\beta)\, e^{-\frac{\beta}{\alpha}\left(\frac{\tau}{\tau_0}\right)^{\alpha}} \left(\frac{\tau}{\tau_0}\right)^{\beta} \tag{1}$$

with the normalizing factor

$$N_{GG}(\alpha,\beta) = \left(\frac{\beta}{\alpha}\right)^{\beta/\alpha} \frac{\alpha}{\Gamma\left(\frac{\beta}{\alpha}\right)} \tag{2}$$

The mean correlation times is given by

$$\langle\tau\rangle = \tau_0 \left(\frac{\alpha}{\beta}\right)^{1/\alpha} \frac{\Gamma\left(\frac{\beta+1}{\alpha}\right)}{\Gamma\left(\frac{\beta}{\alpha}\right)} \tag{3}$$

The maximum of the distribution is located at $\tau = \tau_o$. An example of a GG distribution is presented in Fig. 2. The width parameters α and β can assume values $0 < \alpha,\beta < \infty$. The parameter α specifies the distribution G(lnτ) at large correlation times, explicitly at high $\ln(\tau/\tau_o)$, whereas β fixes the behavior of G(lnτ) at small $\ln(\tau/\tau_o)$. This is easily recognized when the leading terms of G(lnτ) are inspected in the respective limit (cf. also Fig. 2)

$$G_{GG}(\ln\tau) \propto e^{-\frac{\beta}{\alpha}\left(\frac{\tau}{\tau_0}\right)^{\alpha}} \qquad \text{for } \ln(\tau/\tau_o) \gg 0$$

$$G_{GG}(\ln\tau) \propto \left(\frac{\tau}{\tau_0}\right)^{\beta} \qquad \text{for } \ln(\tau/\tau_o) \ll 0 \tag{4}$$

Given the distribution G(lnτ) we obtain the normalized relaxation function Φ(t) via

$$\Phi(t) = \int_{-\infty}^{\infty} G(\ln\tau) e^{-t/\tau} d\ln\tau \tag{5}$$

For the purpose of the present paper we are interested to discuss the complex dielectric permittivity $\varepsilon(\omega)$ which is found by

$$(\varepsilon(\omega) - \varepsilon_\infty)/\Delta\varepsilon = \int_{-\infty}^{\infty} G(\ln\tau) \frac{1}{1+i\omega\tau} d\ln\tau \tag{6}$$

where $\varepsilon_\infty$ is the permittivity at optical frequencies.

The GG distribution, Eq. 1, contains the distribution for a Debye susceptibility as a limit, as $\beta$ goes to infinity. If $0 < \beta < 1$ holds there appears a power law at high frequencies ($\omega\tau_o >> 1$), explicitly $\varepsilon''(\omega) \propto \omega^{-\beta}$. In the case $\beta > 1$ the high frequency exponent of the susceptibility is always –1, the width of the loss peak however may differ significantly. To additionally vary the peak width with the high frequency exponent being < 1 one can adjust the parameter $\alpha$. This gives sufficient freedom to reproduce the salient features of various susceptibilities. At low frequencies ($\omega\tau_o << 1$) the susceptibility depends linearly on frequency, explicitly $\varepsilon''(\omega) \propto \omega$, i.e., it always exhibits the physically correct low frequency limit in contrast to the HN function, for example. As discussed in the appendix the GG distribution well interpolates a KWW susceptibility and in fair approximation also a Cole-Davidson susceptibility. In both cases a linear relationship among the exponent $\beta$ and $\beta_{KWW}$ respectively $\beta_{CD}$ is found in a large parameter interval.

2.2.) Distribution for the $\alpha$-process including the high frequency wing

In many glass formers the susceptibility is described by a single relaxation peak which can be described by the GG distribution, however, at very high frequencies a crossover to another power law is recognized in many systems provided that no further relaxation process obscures this feature (type A systems). Fig. 3 gives an example for the case of glycerol. Note, that inspecting the spectral feature of type A glass formers such as glycerol at $\omega\tau_o >> 1$, only one power law regime can be identified, being located at highest frequencies. This is most easily recognized when the derivative $d\lg\varepsilon''(\omega)/d\lg\omega$ is considered (cf. Fig. 3). Here it is seen that the wing gets less pronounced and might even disappear at high temperatures.

In order to include the high frequency tail into a distribution function of the $\alpha$–process the generalized gamma distribution, Eq. (1), can be extended in the following way [17]

$$G(\ln\tau) = N_{GGE}(\alpha,\beta,\gamma)\, e^{-\frac{\beta}{\alpha}\left(\frac{\tau}{\tau_0}\right)^\alpha} \left(\frac{\tau}{\tau_0}\right)^\beta \left(1 + \left(\frac{\tau\,\sigma}{\tau_o}\right)^{\gamma-\beta}\right) \tag{7}$$

with the normalizing factor

$$N_{GGE}(\alpha,\beta,\gamma)=\alpha\left(\frac{\beta}{\alpha}\right)^{\beta/\alpha}\left(\Gamma\left(\frac{\beta}{\alpha}\right)+\sigma^{\gamma-\beta}\left(\frac{\alpha}{\beta}\right)^{\frac{\gamma-\beta}{\alpha}}\Gamma\left(\frac{\gamma}{\alpha}\right)\right)^{-1} \tag{8}$$

and the mean time constant

$$\langle\tau\rangle=\tau_0\left(\frac{\alpha}{\beta}\right)^{1/\alpha}\frac{\Gamma\left(\frac{\beta+1}{\alpha}\right)+\sigma^{\gamma-\beta}\left(\frac{\alpha}{\beta}\right)^{\frac{\gamma-\beta}{\alpha}}\Gamma\left(\frac{\gamma+1}{\alpha}\right)}{\Gamma\left(\frac{\beta}{\alpha}\right)+\sigma^{\gamma-\beta}\left(\frac{\alpha}{\beta}\right)^{\frac{\gamma-\beta}{\alpha}}\Gamma\left(\frac{\gamma}{\alpha}\right)} \tag{9}$$

Two additional parameters, $\sigma$ and $\gamma$, appear which define the onset of the wing and its exponent $\gamma$, respectively. Again, Fig. 2 displays an example for this extended GG distribution which is abbreviated as GGE. Since usually $\gamma < \beta$ holds Eq. (9) can be approximated by Eq. (3) provided that $\sigma >>1$. In Fig. 3 the fits of the glycerol data by Eq. (7) are included demonstrating a very satisfying interpolation even when the derivative $\mathrm{dlg}\varepsilon''(\omega)/\mathrm{dlg}\nu$ is inspected.

Of course, the GGE distribution allows equally well to interpolate data in the time domain. In Fig. 4a our time domain data of glycerol at T = 184 K are fitted by applying the GGE distribution, in Fig 4b the Fourier transform of both the fit and the experimental data are shown. Clearly, both agree very well and an excess wing can be recognized. For comparison, we include the fit with the GG function which does not take account of the power law contribution at short times respectively high frequencies. As seen from Fig. 4a about 10% of the signal have to be attributed to the excess wing. This example demonstrates that also in the time domain a well adapted fit function reveals more information than applying a simple KWW decay. Actually, by pure visual inspection the excess wing contribution is not easily identified in the time domain data. This is overcome if the derivative of the relaxation function $\phi(t)$ is plotted. In this representation the excess wing is clearly seen as a power law regime $t^{\gamma-1}$. We note that recently other time domain experiments have been carried out by applying the optical Kerr effect [33], and the GGE distribution appears well suited to analyze those data, too.

As there are indications that the high frequency wing disappears in viscous liquids at high temperatures [16] it is important to discuss the conditions for which Eq. (7) (GGE) transforms into Eq. (1) (GG). This limit is reached exactly for $\sigma \rightarrow 0$ or for $\gamma \rightarrow \beta$. However, due to the properties of the Laplace transform there exists a limit with $\sigma > 0$ and $\gamma < \beta$ for which a simple peak susceptibility (without separate high frequency wing) results in $\varepsilon''(\omega)$ although the distribution GGE at $\tau/\tau_o << 1$ still consists of two power law contributions with exponents $\gamma$ and $\beta$, respectively. Indeed, as shown in the appendix the GGE distribution provides an excellent fit of a CD susceptibility with $\gamma = \beta_{CD}$, when a certain parameter constellation of $\sigma(\alpha,\beta,\gamma)$ and $\beta(\gamma)$ (with $\sigma \neq 0$ and $\beta \neq \gamma$) is fulfilled and the parameter $\alpha$ is kept constant at a large value (here $\alpha = 20$) as is usually the case for type A glass formers. The GGE distribution compared to the CD distribution are displayed in Fig. 5 for the condition that the GGE provides the best interpolation of the CD susceptibility (cf. also Fig. A3).

Of course, modeling the CD distribution by means of the GGE function only provides an example of how a simple peak structure in $\varepsilon''(\omega)$ can be reproduced although neither of the limits $\sigma = 0$ and $\gamma = \beta$ is reached. It appears, however, that the short time tail of the CD distribution due to its particular shape can provide a reliable estimate for the condition under which a separate high frequency wing disappears in $\varepsilon''(\omega)$ in a more general way. For this purpose, a function $\sigma_c(\alpha,\beta,\gamma)$ can be calculated such, that the absolute short-time asymptote of the GGE and the CD distribution become identical, explicitly, assuming that

$$\lim_{\tau\to 0} G_{GGE}(\ln\tau) \quad \to \quad \frac{\sin(\pi\gamma)}{\pi}\left(\frac{\tau}{\tau_o}\right)^{\gamma} \tag{10}$$

we find the constraint

$$\sigma_c(\alpha,\beta,\gamma) \;=\; \left(\frac{\left(\frac{\beta}{\alpha}\right)^{\frac{\beta}{\alpha}}\left(\frac{\alpha\pi}{\sin(\pi\gamma)} - \left(\frac{\alpha}{\beta}\right)^{\frac{\gamma}{\alpha}}\Gamma(\gamma/\alpha)\right)}{\Gamma(\beta/\alpha)}\right)^{\frac{1}{\beta-\gamma}} \tag{11}$$

Thus, irrespective of the particular shape of the α–peak itself, $\sigma_c$ assures that no wing is present in $\varepsilon''(\omega)$. This holds provided that the α–peak is similar to a CD line shape, which turns out to hold for $\alpha \geq 1$. As is discussed in the appendix the constraint, Eq. (11), yields a σ as a function of β which is very close to that obtained for a free fit of a CD susceptibility. Note that a free fit of a CD susceptibility by the GGE distribution always leads to $\gamma = \beta_{CD}$. Concluding, instead of a free fit of the data the constraint can be applied to guarantee that the GGE is well reproducing a simple peak susceptibility which is actually very close to the CD limit, or vice versa, the constraint may be applied in order to check how close an actual susceptibility function is to a simple peak susceptibility. In this case the fitted parameter σ is compared with $\sigma_c(\alpha,\beta,\gamma)$. The latter approach will be applied in section 3.1.

Having thus introduced the GGE distribution the two limiting cases, namely the CD limit and the strong wing limit can be reproduced within a single distribution function. This allows to monitor a possible emergence of a wing contribution while cooling. Indeed, as we will demonstrate in section 3.1 the salient features of the slow dynamics in type A glass formers are reproduced by applying solely this distribution function.

### 2.3 Distribution for the β-process

As discussed, several glass formers exhibit a well detectable secondary relaxation which may vary significantly in its strength. This β–process persists in the glass where its time constant is well described by an Arrhenius law. Thus, there are good reasons to assume that this process is less collective as compared to the α–process and may be described by a distribution of thermally activated processes below $T_g$. This implies that the parameters of the distribution $G_\beta(\ln\tau)$ exhibit a certain temperature dependence which is compatible with an underlying temperature independent distribution of activation energies. However, as discussed e.g. by Boettcher and Bordewijk [29], this property does not apply to many commonly used susceptibility functions, e.g., the Cole-Davidson (CD), Cole-Cole (CC) or Havriliak-Negami (HN) function. For example, one can demonstrate that if a thermally activated process at some temperature T has the shape of, e.g., a HN function, then at any other temperature the process is no longer compatible with a HN line shape. Therefore, a suitable distribution $G_\beta(\ln\tau)$ requires that for $\tau(T) = \tau_o \exp(E/RT)$, the width parameters (e.g. $\alpha(T)$ or $\beta(T)$ in the case of the GG distribution) can be chosen as a function of temperature such that, if one turns $G_\beta(\ln\tau)$ into a distribution of activation energies g(E), the latter becomes temperature independent:

$$G_\beta(\ln\tau(T), a(T), b(T))\, d\ln\tau(T) = g(E)\, dE \neq f(T) \tag{12}$$

It is easily shown that indeed the GG distribution, Eq. (1), meets theses requirements and thus seems equally well suitable to describe the β–process. However, as this distribution always yields a Debye-like low frequency slope in $\varepsilon''(\omega)$, often the interpolation of experimental data in the region of the minimum between α– and β–process turns out to be rather poor.

The following distribution is more suitable for secondary processes:

$$G_\beta(\ln\tau) = N_\beta(a,b)\frac{1}{b\left(\frac{\tau}{\tau_m}\right)^{a} + \left(\frac{\tau}{\tau_m}\right)^{-ab}} \tag{13}$$

with $$N_\beta(\alpha,\beta) = \frac{a(1+b)}{\pi}\, b^{b/(1+b)} \sin\left(\frac{\pi b}{1+b}\right) \tag{14}$$

being the normalizing factor. $\tau_m$ defines the maximum of the distribution. Note that Eq. (13) is not to be confounded with the Jonscher description, which is defined as a function in the frequency domain [29]. The corresponding permittivity bears resemblance with the HN function. The parameter *a* controls a symmetric broadening of the peak whereas *b* takes account of an additional asymmetry of the high frequency slope. For $a < 1$ and $ab < 1$, both sides of the peak show power laws $\omega^a$ and $\omega^{-ab}$, respectively. Fig. 2 gives an example of $G_\beta(\ln\tau)$. With *a*, $ab > 1$ there appear power laws ω resp. $\omega^{-1}$, with the line shape approaching the Debye function for large values of *a* and *ab*. A physically reasonable long-time behavior of the overall relaxation function Φ(t) including α– and β–relaxation is ensured when this distribution is applied within the Williams-Watts approach (cf. below). Assuming, $\tau(T) = \tau_o \exp(E/RT)$, and correspondingly, $\tau_m(T) = \tau_o \exp(E_m/RT)$, one obtains a temperature independent distribution of activation energies exactly if $a \propto T$ and $b$ = const. Putting $a = c/R\,T$ one comes to

$$g_\beta(E) = N_\beta(c,b)\frac{1}{be^{c(E-E_m)} + e^{-cb(E-E_m)}} \tag{15}$$

Fig. 6 gives an example of interpolating the β–process in an epoxy compound ($T < T_g$) [33,34]. In this case, obviously the susceptibility is not symmetric as is recognized if one compares it with that derived from a Gaussian distribution. Further applications of Eq. (13) will be given in section 3.2.

## 2.4 The Williams-Watts approach

The Williams-Watts (WW) approach has been applied to describe both α- and β–process [17,30,35,36]. This ansatz also allows to describe the merging of both processes above $T_g$. The WW approach assumes that due to the β-process a partial relaxation occurs at short times which at later times is completed by the α–process. Moreover, both processes are assumed to be statistically independent. Consequently, the corresponding relaxation functions have to be multiplied:

$$\Phi(t) = \lfloor (1 - S(T)\Phi_\beta(t) + S(T) \rfloor \Phi_\alpha(t) \qquad (16)$$

with $\Phi_\alpha$(t) and $\Phi_\beta$(T) representing the normalized relaxation function of α– and β–process, respectively, and S(T) stands for the relative relaxation strength of the α-process. Formally, S(T) has a similar role as the non-ergodicity parameter f(T) discussed within the mode coupling theory though the latter refers to the partition of slow dynamics and fast dynamics (ν > 1 GHz) which is not considered in the present contribution. Here the WW approach will be applied provided that a secondary relaxation peak can be identified.

A technical aspect is worthwhile to be mentioned here. In order to obtain the susceptibility function from the relaxation function modeled within the WW approach a Fourier transformation has to be carried out. As the relaxation function extends over many decades in time the FFT algorithm, which increments time linearly, is not adequate and we rather apply a variant of the Filon algorithm which allows to increment time on a logarithmic scale [37].

We note that WW ansatz can also be taken to describe the susceptibility of type A glass formers if one assumes that the excess wing is a spectral feature originating from a separate relaxation process and the latter having a time constant quite close to that of the α-process. Using both the GG and $G_\beta$ distribution with certain parameter constellation a very satisfying interpolation is obtained. For example, the conditions $\tau_m = \tau_0$, $b = 1$ and $a = \gamma$, and $1 - S = \sigma^{\gamma-\beta} N_{GGE}/N_\beta$ may be chosen; the latter condition is obtained by assuming that the GGE and $G_\beta$ distribution exhibit the same dependence at short τ, a property easily fulfilled since in this limit both distribution functions are described by simple power laws. Having introduced with Eq. (13) a distribution exhibiting power law behavior at low and high frequencies, this approach is particularly appropriate for describing systems where the β–process continuously degenerates to an excess wing contribution as is the case in binary glass formers [32] and was also observed for a homologous series of polyalcohols [38].

# 3 Some applications

## 3.1 Type A glass formers

Since long the main relaxation in glass formers attracted much attention as one hoped to discover some universal laws which determine the evolution of the shape of the dynamic susceptibility. A starting point was the frequency temperature superposition principle (FTS) which states that the shape of the main response is independent of temperature. However, due to technical progress it became obvious that additional relaxational features emerge while supercooling which have to be taken into account. In the present section we will restrict our discussions to systems showing a pronounced excess wing and where no secondary relaxation peak is recognized (type A glass formers).

Analyzing the dielectric loss of type A glass formers we apply the GGE distribution, Eq. (7). By doing so we first of all aim at simply parameterizing the experimental data without implying any physical interpretation of the nature of the wing contribution. In Fig. 7 we present dielectric loss data for two glass formers (in addition to that for glycerol in Fig. 3) which obviously do not exhibit any secondary relaxation peak, namely propylene carbonate (PC) and 2-picoline. We note that in the case of glycerol our previous data [17] have been extended to very low frequencies by applying time domain measurements [32]. As in the case of glycerol an almost perfect interpolation is provided by applying free fits with the GGE distribution for PC and picoline.

In all cases very similar time constants $\tau_\alpha$ are found as reported in the literature [2,8,17]. As expected the wing contribution does not alter the average time constant as compared to that applying, e.g., a simple CD distribution. The high frequency contribution of the GGE susceptibility is described by the exponent $\gamma$ and by its onset $\sigma$. In order to show the change of these parameters while cooling, we have plotted them in Fig. 8 as a function of the time constant of the $\alpha$–process, $\lg\tau_\alpha$. This allows one to compare the evolution of the suscepti-bility independent of the fragility of the glass former. Clearly, the data of $\sigma$ (cf. Fig. 8c) and of $\gamma$ (cf. Fig. 8b) exhibit a universal behavior in the entire range studied. The onset parameter $\lg\sigma$ linearly increases with $\lg\tau_\alpha$ while the exponent $\gamma$ decreases continuously. Concerning the two parameters, $\alpha$ and $\beta$, which describe the $\alpha$–peak, the parameter $\alpha$ may be kept constant for all temperatures for a given glass former but its value varies among the different systems (cf. table 1). The parameter $\beta$ of the three systems (cf. Fig. 8a) is almost identical in the range $-3 < \lg\tau_\alpha$, and slowly increases while the structural relaxation accelerates. At $\lg\tau_\alpha < -3$ the parameter $\beta$ increases stronger and somewhat differently for the systems studied.

Although these results were obtained applying a certain distributions $G(\ln\tau)$ they are essentially independent of the particular choice of a model function. This is most directly seen when the different data sets are scaled by temperature independent vertical shifts in order to provide the best coincidence in the high frequency region. Indeed, as expected from the fit results, very similar high frequency contributions are observed in Fig. 10 whereas the $\alpha$–relaxation peak itself is different, what is reflected in different but constant values of the parameter $\alpha$ (cf. table 1). At least in these systems the mesoscopic dynamics is rather universal in the sense that the wing contribution manifests itself in a very similar way whereas the slowest dynamics differs along different $\alpha$ parameters. We emphasize that these findings clearly disprove the claim of the Nagel scaling. The latter implies a universal relationship between the exponent $\gamma$ and a parameter defining the width of the $\alpha$–relaxation peak. This is clearly not observed as is disclosed, e.g., by the data in Fig. 10.

How are these results to be interpreted? The three systems investigated show the same behavior independent of their fragility. In all cases the onset parameter $\sigma$ approaches the order of 1 and in this regime ($\lg(\tau_\alpha/s) < -3$) the parameter $\beta$ strongly increases. According to the discussion in the appendix this finding signals that the susceptibility function resembles more and more a CD function with a high frequency exponent defined by $\gamma = \beta_{CD}$, i.e., the high frequency wing essentially disappears at high temperatures.

In order to apply a more precise criterion to find the condition where actually the susceptibility can be described by a simple peak without a wing contribution we apply Eq. (11) which, as discussed in section 2.2, is a consequence of implying the constraint, Eq. (10), on the onset parameter $\sigma$. From each set of parameters $\beta$ and $\gamma$ obtained by the free fit for a given temperature we have calculated the parameter $\sigma_c$ which fulfills Eq. (11). The result is included in Fig. 8c. From this it is clearly seen that the calculated $\sigma_c$, on the one hand, and the actually fitted parameter $\sigma$, on the other, converge in the range $-8 > \lg\tau_\alpha > -9$ indicating that the wing indeed disappears in this range, and, taking into account that the parameter $\beta$ strongly increases the CD limit is reached in good approximation. In other words, in the fluid regime ($\lg\tau_\alpha < -8$) the susceptibility is described by a CD function whereas at low temperatures, i.e., at $\lg\tau_\alpha > -8$, successively the wing contribution appears.

Finally, we can ask whether the frequency-temperature superposition principle (FTS) is rediscovered in our data analysis. Since the excess wing contribution changes with temperature only the shape of the $\alpha$–peak itself may be subject to the FTS. This shape is determined by the parameters $\alpha$ and $\beta$. According to Eq. (4) the $\alpha$ parameter defines the behavior of the GGE distribution at longest times and actually it is chosen constant for a given glass former. Thus, regarding the slowest response of the system, the FTS holds for all temperatures. Concerning the parameter $\beta$ which specifies the GGE distribution at intermediate times and the $\alpha$–peak at its high frequency side it appears to be rather temperature independent in the range $\lg\tau_\alpha > -3$, i.e., at low temperatures. At higher temperatures where the wing and $\alpha$–peak contribution approach each other any statement becomes difficult.

We have measured another glass former, namely m-tricresyl phosphate (m-TCP) for which the data do not allow an unambiguous determination of the parameters of the wing contribution as the analyses is hampered by the appearance of a shoulder at the high frequency side of the $\alpha$–peak [15]. Yet, the low temperature data are still compatible with the aforementioned universal behavior. This is demonstrated in Fig. 9 where the data of m-TCP have been included in a way to provide the best coincidence at highest frequencies. The same holds for the data of ethylene glycol [39] which have also been added.

### 3.2 Type B systems

In this section we apply either the GG distribution, Eq. (1) or the GGE distribution, Eq. (7) for the $\alpha$–process together with the WW approach, Eq. (16), and the $G_\beta$ distribution for the $\beta$–process in order to describe the dynamic susceptibility in type B glass formers, *i.e.*, in systems where a secondary relaxation is clearly identified. Of course, such a fit needs data sets extending over many decades in frequency. We start with analyzing the data for 3-fluoro aniline (FAN) for which we have extended our previous measurements [32] now also including time domain data, and 12 decades in frequency are covered. The data are presented in Fig. 10a. As is clearly seen, in addition to a secondary relaxation peak also a excess wing contribution shows up. Up to now it is not clear whether the frequency range where the wing contribution actually appears in the data, suffices to determine the wing parameters unambiguously. Nevertheless, we have included the line shape parameters obtained from the fits in Fig. 8. Whereas the parameter β follows the behavior of the type A systems the exponent γ and the onset parameter σ show quantitatively a somewhat different dependence on $\lg\tau_\alpha$. However, qualitatively the same behavior is observed in the sense that the exponent γ increases strongly above $T_g$ and that at $\lg\tau_\alpha \cong -8$ the wing contribution disappears on heating. Concerning the width parameter *a* of the β–process the behavior expected for a thermally activated process is found, *i.e.*, $a^{-1}$ depends linearly on 1/T as is shown in Fig. 11a. We emphasize that the parameter $a^{-1}$ becomes zero at a finite temperature $T_\delta$. This may be explained within the frame of the Meyer-Neldel rule [17]. The relaxation strength shows the temperature dependence known for the β–process [17]. It is essentially constant in the glass ($T<T_g$), whereas a strong increase is observed above $T_g$ (cf. Fig. 11b).

Next, we discuss the glass former toluene [17,40]. Fig. 10b displays the data. In this case, no wing contribution is observed in the data. Thus the α–relaxation peak can be interpolated by the GG distribution, Eq. (1). Again, Eq. (13) for the β–process together with the WW approach is applied. Clearly, a very satisfying fit is found. For toluene the width parameter α and β strongly differ from all reported so far in Fig. 8. The α–relaxation peak is significantly broader as compared to that of type A systems. The relaxation strength and width parameter for the β–process behave as expected (cf. Fig. 11b).

Finally, we want to point out an application of the introduced distribution functions for a binary system. This class of glass formers exhibits very broad dielectric loss spectra, in particular, significantly broader than in neat substances [32,41]. In addition, a strong secondary relaxation process may show up if the concentration is chosen properly although the corresponding neat systems do not show a β-process. Explicitly, we consider 2-picoline in o-terphenyl [32]. Fig. 12 presents the data and the fits. Again, a very satisfying interpolation is provided. For a full analysis and discussion of the parameters cf. ref. [32].

## 4. Discussion and conclusions

The generalized gamma (GG) distribution and an extension thereof (GGE) have been discussed for interpolating the dynamic susceptibility in simple glass formers. The GGE distribution is flexible enough to excellently describe both a CD function as well as a α–process susceptibility with a more or less pronounced wing contribution. In addition, we have introduced a distribution for interpolating the β–process spectra which appears to be more appropriate than the commonly applied functions as it is compatible with underlying thermally activated dynamics. The function leads to high and low frequency power laws with exponents smaller than one and, if required, asymmetric spectra can be reproduced. Thus, future investigations can address systematically the question whether the wing contribution can be regarded as a kind of β–process [32].

Applying the GGE distribution to three well investigated type A glass formers we can show that this distribution reproduces the salient features of the susceptibility below say $10^9$Hz.

Strong evidence is compiled that the excess wing contribution disappears at a certain stage of supercooling defined by a characteristic correlation time on the order of $\tau_\alpha \cong 10^{-8}$s. Moreover, we find indications that the α–process spectrum at high temperatures reduces to a CD susceptibility in good approximation. The parameters characterizing the wing contribution for the systems studied are governed by a universal dependence when plotted rather as a function of the time constant $\tau_\alpha$ than as a function of temperature. In contrast to that, the α–peak itself is different for the systems. In other words, the mesoscopic instead of the slow dynamics seems to exhibit universal behavior. Whether this can be confirmed for other type A systems is a matter of future investigations. The wing parameters in type B systems turn out be different from those in type A systems. It appears that the presence of a β–process leads to deviations from the universal line shape pattern characteristic of the type A systems investigated here. Also, the α–peak width increases when a β–process is present, an observation also stressed by Ngai et al. [41].

Here we want to comment on the problem whether the wing is actually a secondary relaxation process of Johari-Goldstein type [21-23]. Clearly, below $T_g$ the excess wing contribution has characteristics of a secondary relaxation process, and also above $T_g$ it is not necessarily to be interpreted as part of the α–process, as the wing contribution exhibits universality to a higher degree than the actual peak does, thus disproving the claim of the Nagel scaling. Recent aging experiments also point into the direction that the excess wing may become a separate process [22]. However, we think that the wing contribution is well distinguished from a JG secondary process since (i) above $T_g$ there are no indication that for describing the excess wing more than a power law contributions is needed. (ii) Often the excess wing is present in both type A and type B systems (cf. FAN data in Fig. 11a and also ref. 43 and 44), and (iii) it is observed in the light scattering susceptibility whereas there are hints that the JG does not show up in these experiments [8,45]. Recent dielectric experiments applying pressure also come to the conclusion that excess wing and JG process are different phenomena [24].

Though the approach presented here is purely phenomenological we think the results presented in particular for type A systems are independent of the interpretation of the wing phenomenon and all in all they provide a clear phenomenological picture. Supercooling a simple glass former a high and a low temperature scenario has to be distinguished. For correlation times shorter than $\tau_\alpha \cong 10^{-8}$s the high temperature scenario comprises a loss spectrum essentially given by CD function describing the α–process spectrum and some fast dynamics contribution (not analyzed here). In our recent light scattering (LS) experiments on toluene and 2-picoline we indeed have proven this scenario to hold well [7,8]. At low temperatures the emerging wing contribution is the sole change of the line shape in type A systems which actually also shows up in the LS spectra below $T_c$ [7,8].

MCT provides a consistent description of the high temperature dynamics in many glass formers. Thus, the physical picture sketched in the present publication suggests that the breakdown of this high temperature scenario actually coincides with the emergence of the wing contribution. Indeed comparing the corresponding temperature associated with the characteristic correlation time at $10^{-8}$ s for which the wing contribution disappears upon heating and the critical temperature $T_c$ defined within an (idealized) MCT analysis high correlation is found for the various glass formers. Table 2 compiles the results reported in the literature. The idea of a “magic” relaxation time has recently been discussed by Sokolov and Novikov [46]. It is also close to the value discussed by Goldstein indicative of a crossover from liquid-like to solid-like motion [47]. Thus, from a phenomenological point of view it is tempting to identify the dynamical crossover with the emergence of the excess wing which is well defined since a simple relation holds for the onset, explicitly, $\lg(\sigma/\sigma_c) = C \lg\tau_\alpha/\tau_c$, with C = 0.26 and $\sigma_c$ is actually on the order of one and $\tau_c$ on the order of $10^{-8}$s (cf. Fig. 8).

Concluding we say that although clarifying the physical nature of the slow response of a glass former is still a future task to be solved the here proposed distribution functions yielding a complete interpolation of the susceptibility spectra provide a clear cut identification of the spectral changes occurring while supercooling a simple glass former, i.e. independent of any theoretically guided analysis clearly a crossover from a high temperature to a low temperature scenario emerges. Provided that this can generally be identified with the critical temperature of MCT, which is determined by analyzing the high temperature data, the approach presented here allows for an identification of this crossover from the low temperature side.

Acknowledgement

Helpful discussions and suggestions with V. N. Novikov are appreciated, and we thank F. Kremer for providing the dielectric data on ethylene glycol.

Table 1: Width parameter $\alpha$ of the GG or GGE distribution

| | | |
|---|---|---|
| glycerol | 10 | (GGE) |
| 2-picoline | 5 | (GGE) |
| propylene carbonate (PC) | 20 | (GGE) |
| fluoro ananiline (FAN) | 5 | (GGE) |
| toluene | 1 | (GG) |

Table 2: Correlation between the temperature $T_{-8}$ at which $\tau_\alpha \cong 10^{-8}$ s and the critical temperature $T_c$ defined within an (idealized) MCT analysis

| system | $T_{-8}$ / K | ref. | $T_c$ / K | ref. |
|---|---|---|---|---|
| calcium potassium nitrate | 400 | 48 | 378 | 5 |
| 2-picoline | 162 | 8 | 162 | 8 |
| propylene carbonate | 200 | 2 | 187 | 3 |
| o-terphenyl | 299 | 49 | 289 | 6 |
| salol | 268 | 50 | 256 | 4 |
| toluene | 150 | 7 | 153 | 7 |

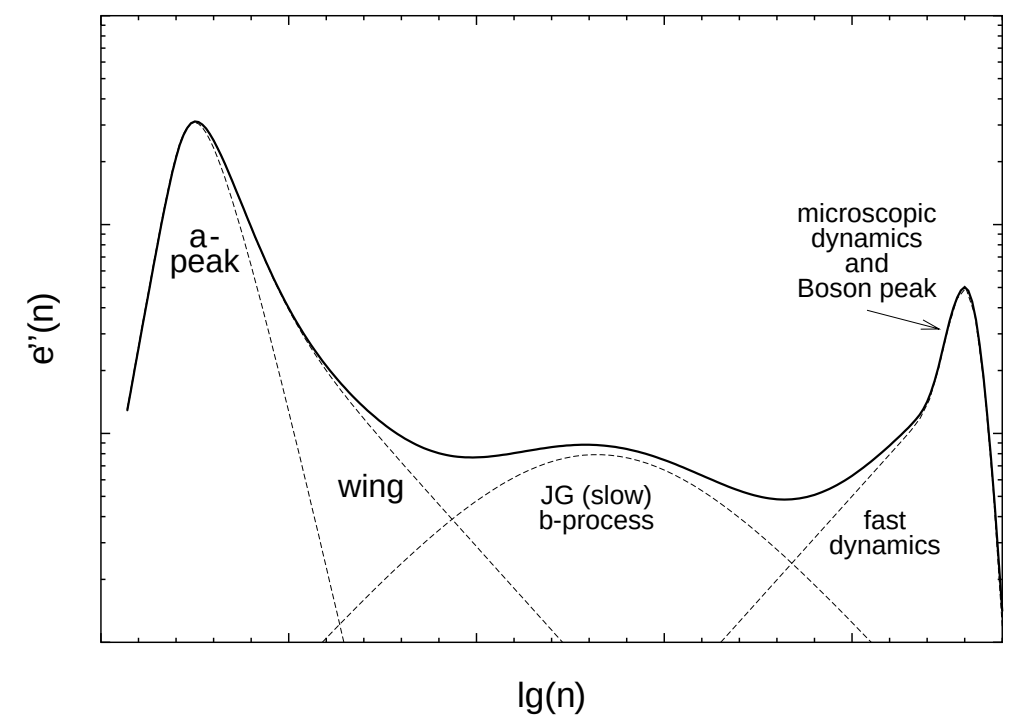

**Figure 1:** Schematic representation of the various relaxation processes that in principle can appear in supercooled liquids.

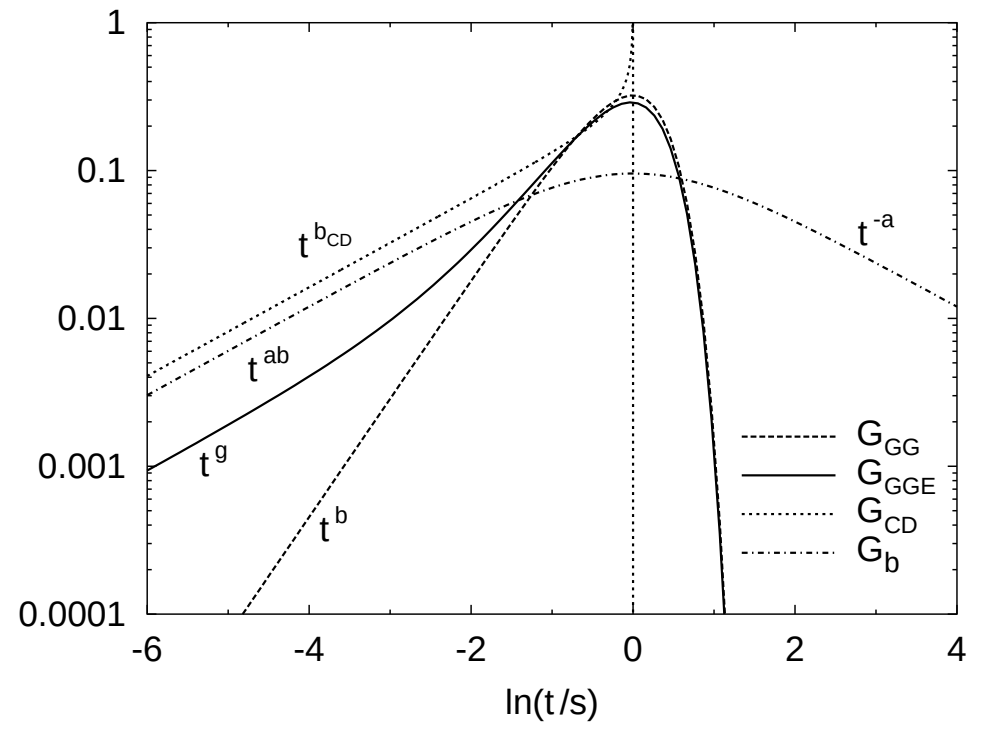

**Figure 2:** Distributions of correlation times discussed in the present paper; generalized gamma function ($G_{GG}$), extended generalized gamma function ($G_{GGE}$), Cole-Davidson distribution ($G_{CD}$) and the distribution introduced for the $\beta$-process ($G_\beta$); power law limits are indicated.

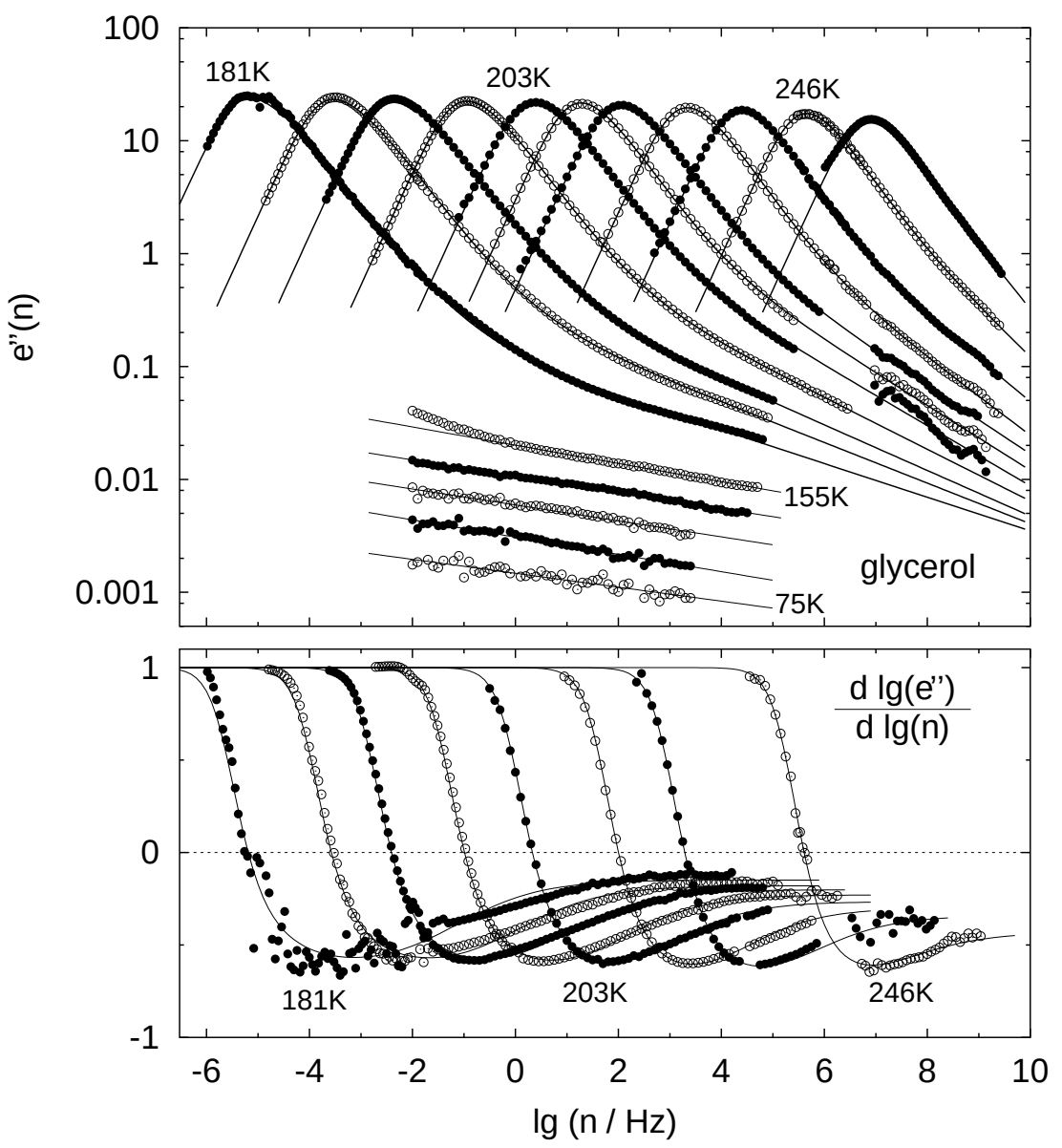

**Figure 3:** (top): Dielectric loss spectra of glycerol interpolated by applying the GGE distribution (solid lines); (bottom): corresponding derivatives, $d\lg\varepsilon''(\nu)/d\lg\nu$, of both data (points) and fit (solid lines); taken from ref. [17].

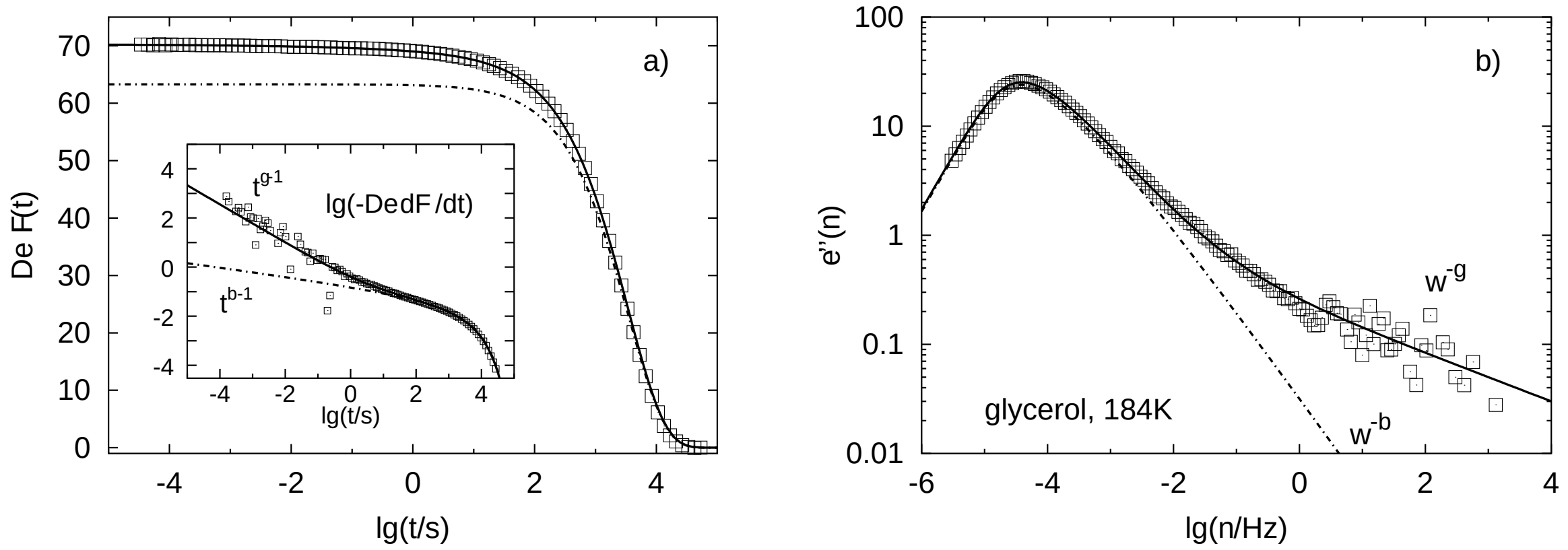

**Figure 4: a.)** Time domain data $\Delta\varepsilon\,\Phi(t)$ of glycerol (squares) fitted by the GGE distribution (solid line); for comparison interpolation by the GG distribution (dash dotted line); **insert:** derivative of both relaxation function $\Phi(t)$ and fit, indicated are the respective power law regimes; **b.)** Fourier transform of both data and fit.

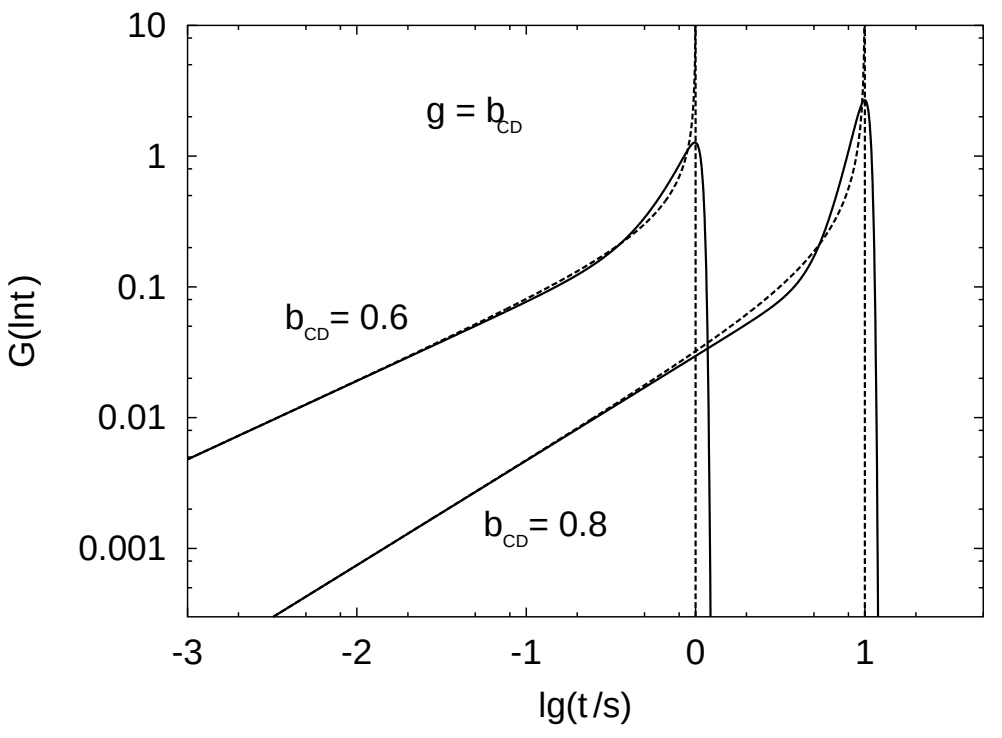

**Figure 5:** Cole-Davidson distribution approximated by the GGE function.

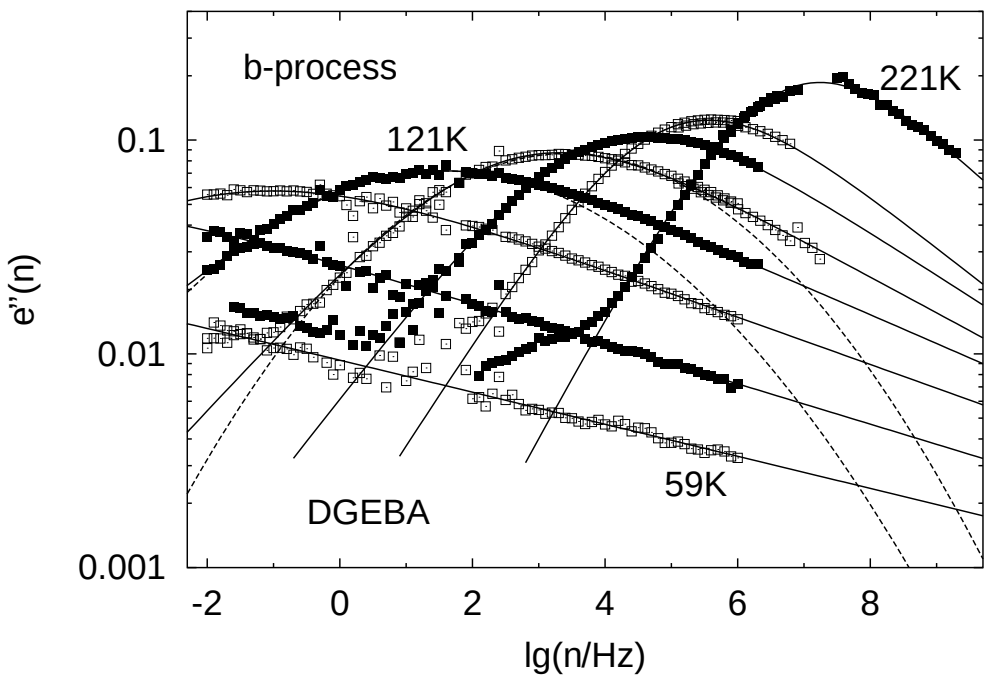

**Figure 6:** Dielectric loss data of the epoxy compound diglycidyl ether of bisphenol A (DGEBA) exhibiting a well separated and asymmetric $\beta$-relaxation peak ($T < T_g = 253\,\mathrm{K}$) interpolated by applying Eq. (13) (solid lines); for comparison fits with a log-Gaussian distribution are shown (dashed lines).

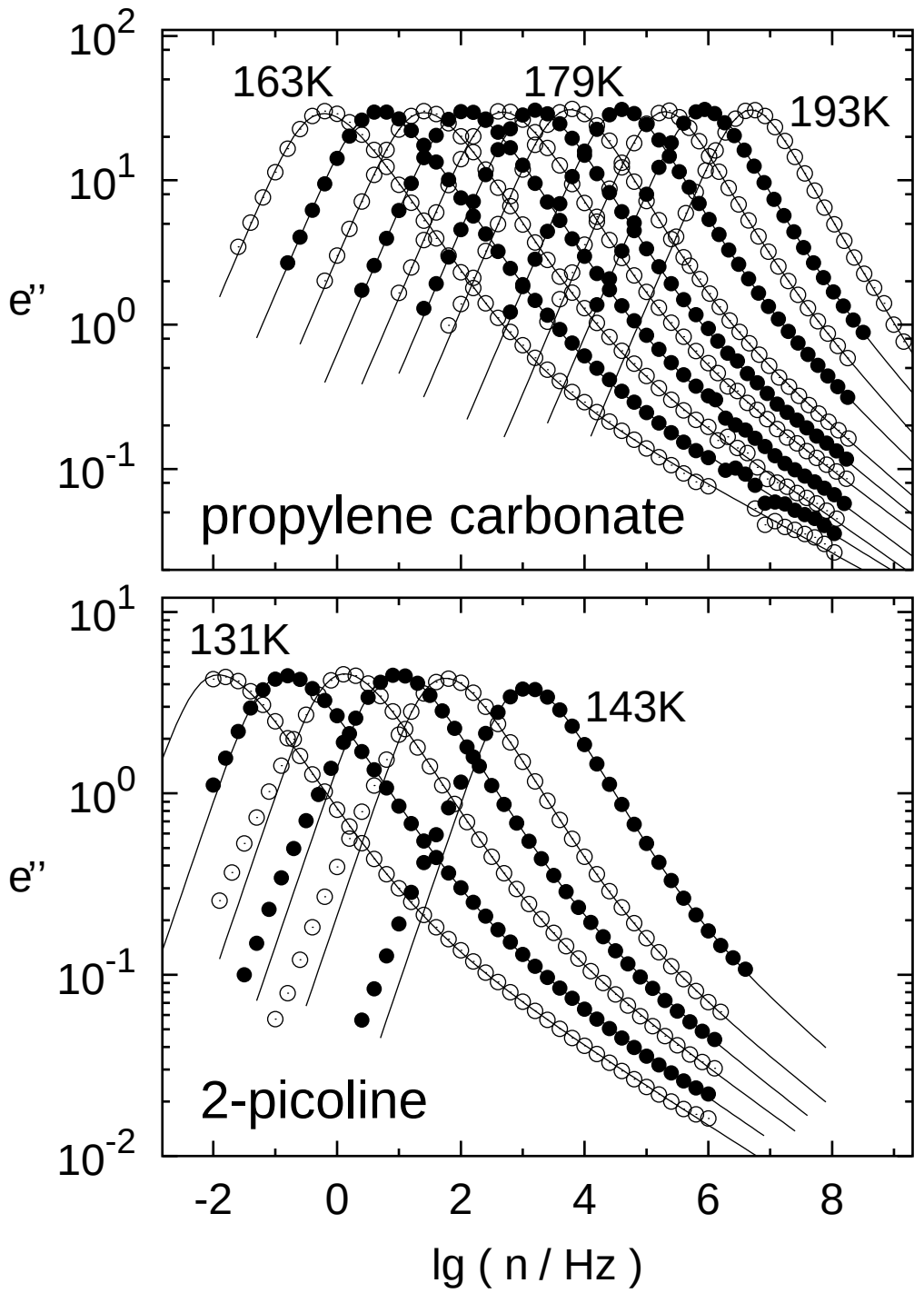

**Figure 7:** Dielectric loss data of propylene carbonate (top) and 2-picoline (bottom); solid lines: interpolation by applying the GGE distribution, Eq. (7).

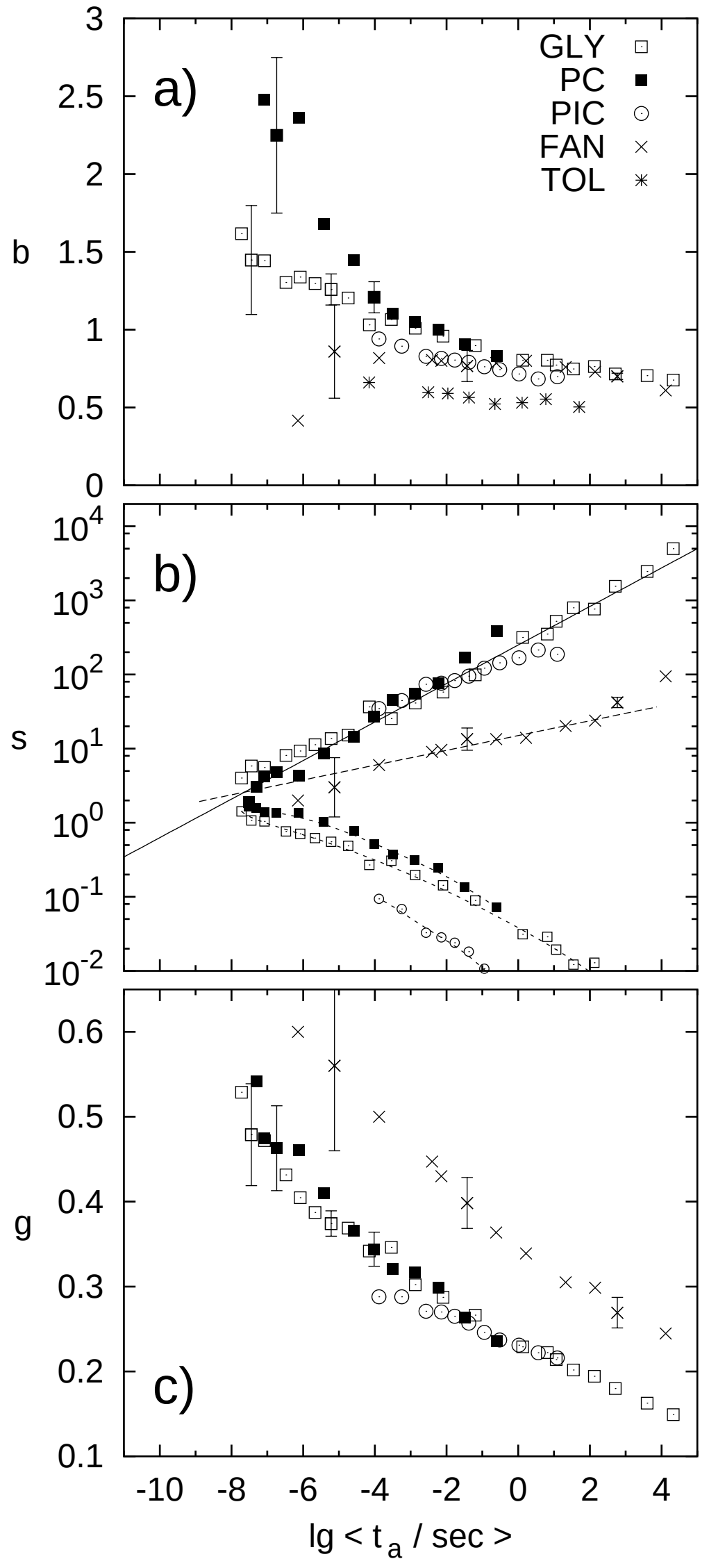

**Figure 8:** Fit parameters $\beta$, $\sigma$, $\gamma$ of the GGE distribution, Eq. (7), applied to the dielectric spectra of type A systems to interpolate the data of glycerol (Fig. 3), propylene carbonate and 2-picoline (Fig. 7) and fluoro aniline (Fig. 10a); in addition the $\beta$-values of toluene (Fig. 10b) are shown; symbols with dashed line represent the constraint $\sigma_c(\alpha, \beta, \gamma)$, Eq. (11), calculated from the parameters of the free fit; solid and dashed straight line: guide for the eye.

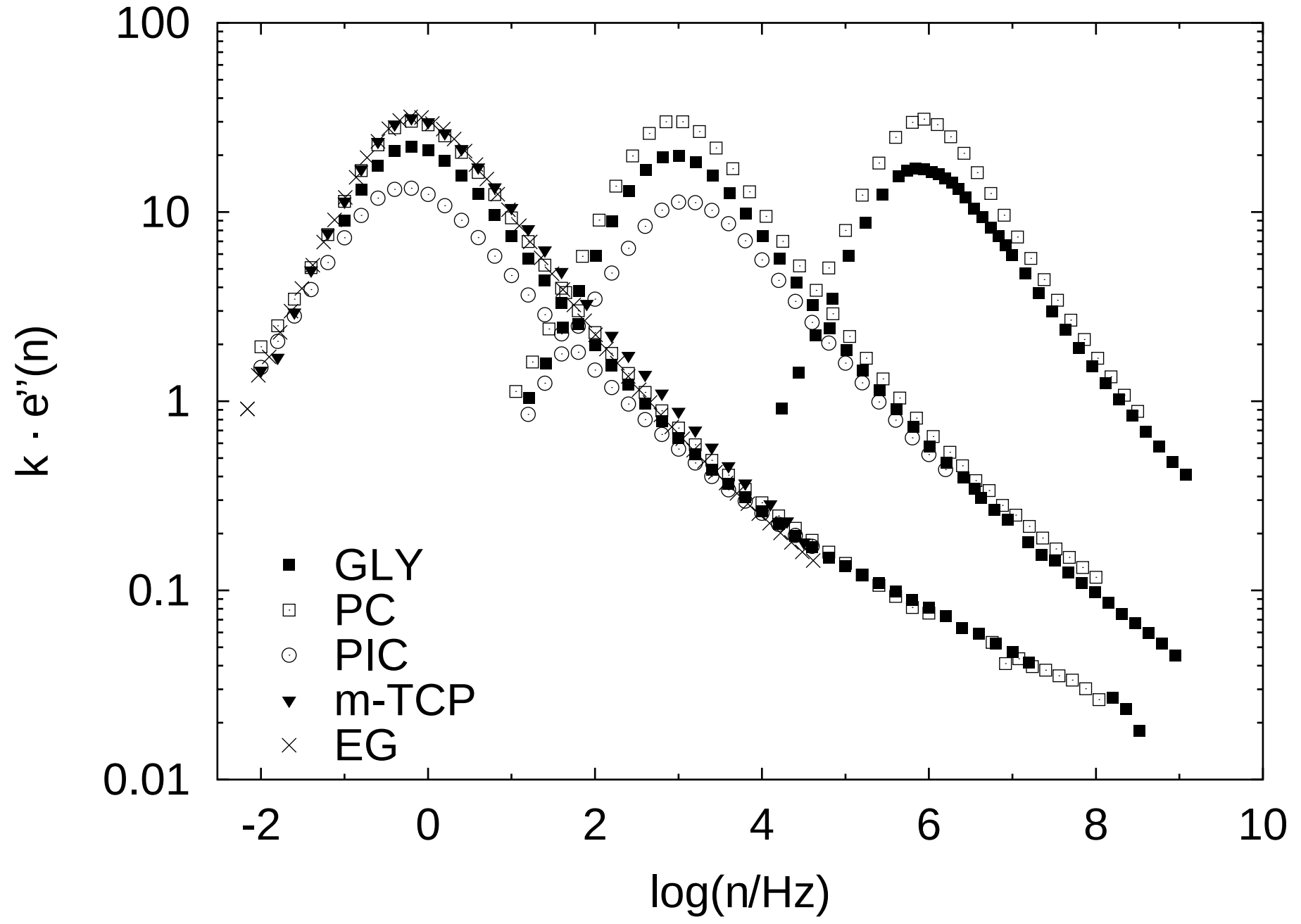

**Figure 9:** Dielectric loss data of type A glass formers shifted by a temperature independent factor k to provide coincidence at highest frequencies: glycerol (GLY, $k = 1$), propylene carbonate (PC, $k = 1$), 2-picoline (PIC, $k = 3$), m-tricresyl phosphate (m-TCP, $k = 15$) and ethylene glycol [37] (EG, $k = 1.4$).

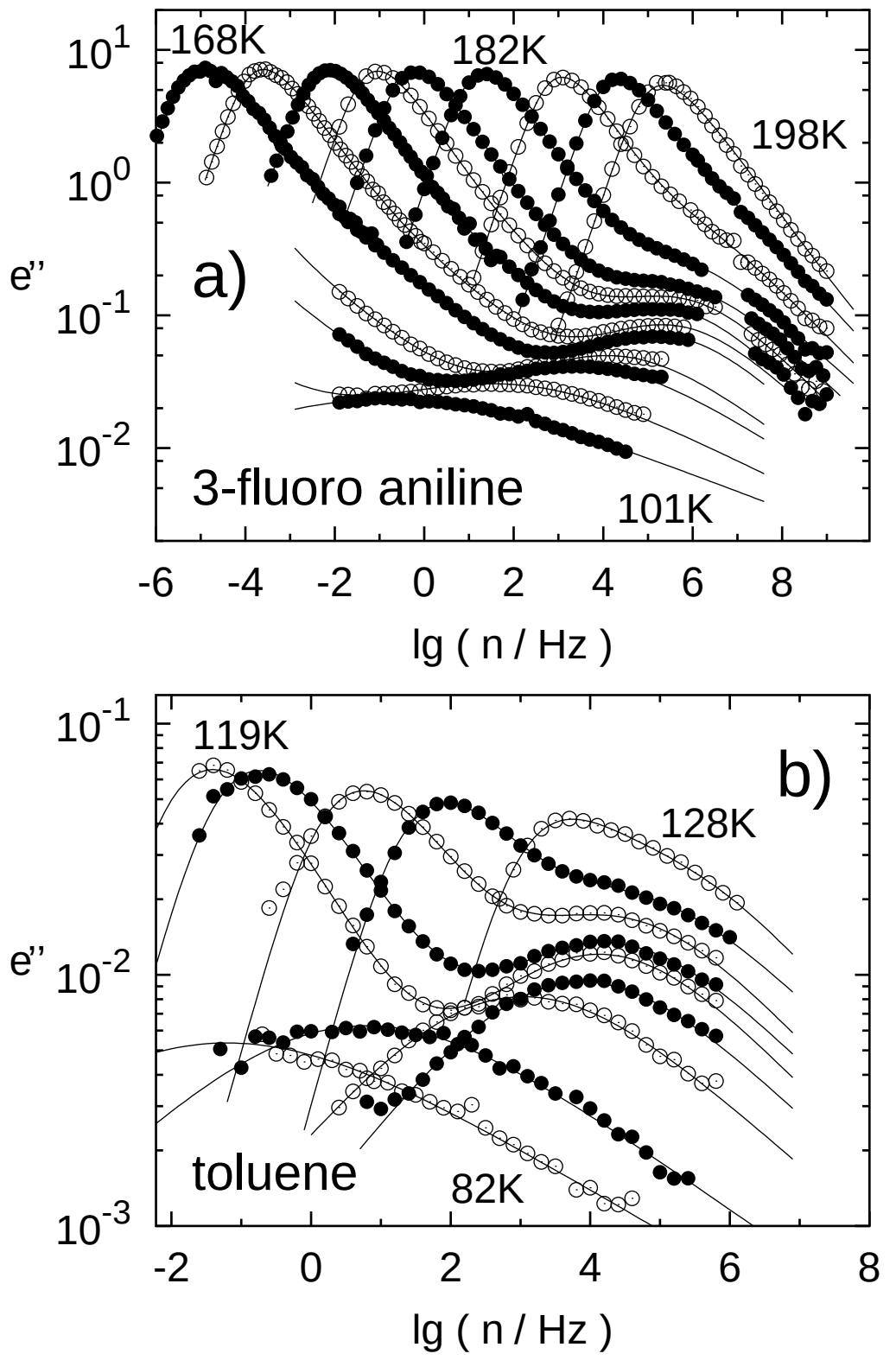

**Figure 10:** Dielectric loss data for type B systems: 3-fluoro aniline and toluene; solid lines: interpolation applying Eq. (7) (3-fluoro aniline) or Eq. (1) (toluene) together with Eq. (13) within the frame of the Williams-Watts approach, Eq. (16).

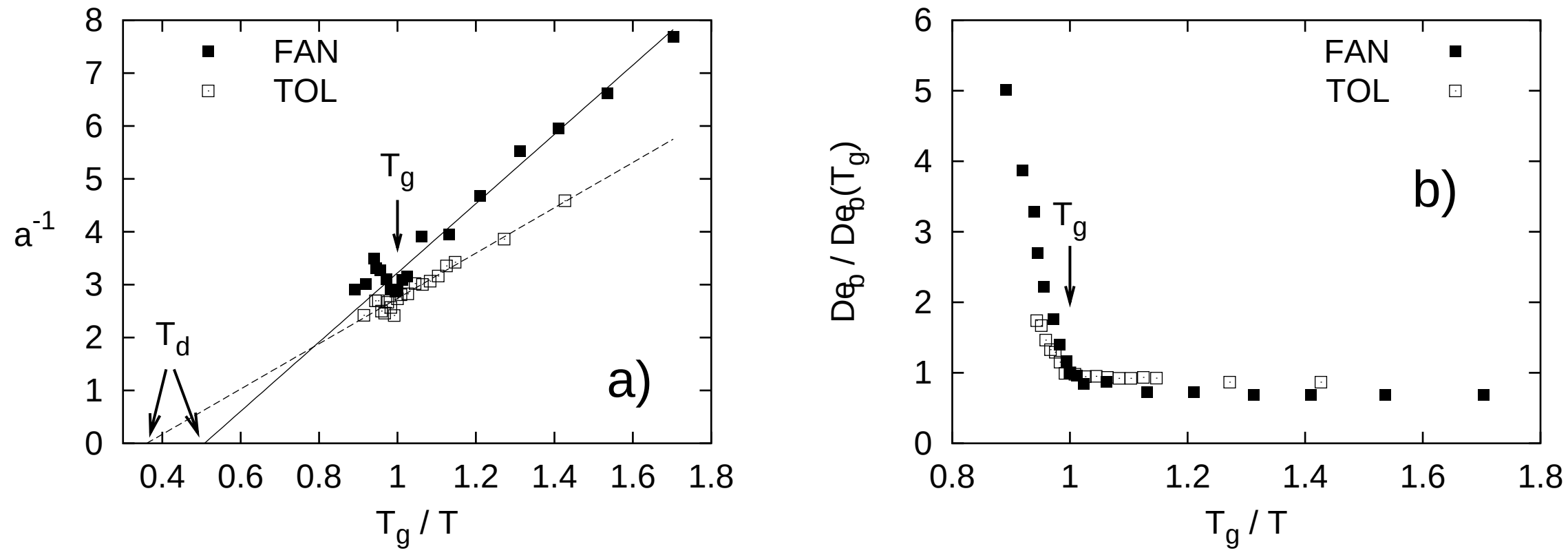

**Figure 11:** Parameters characterizing the $\beta$-process in fluoro aniline (FAN) and toluene (TOL): **a.)** width parameter $a^{-1}$ as a function of reduced reciprocal temperature $T_g/T$; solid and dashed line: behavior expected for a thermally activated process; **b.)** corresponding relaxation strength $\Delta\varepsilon_\beta$ as function of $T_g/T$.

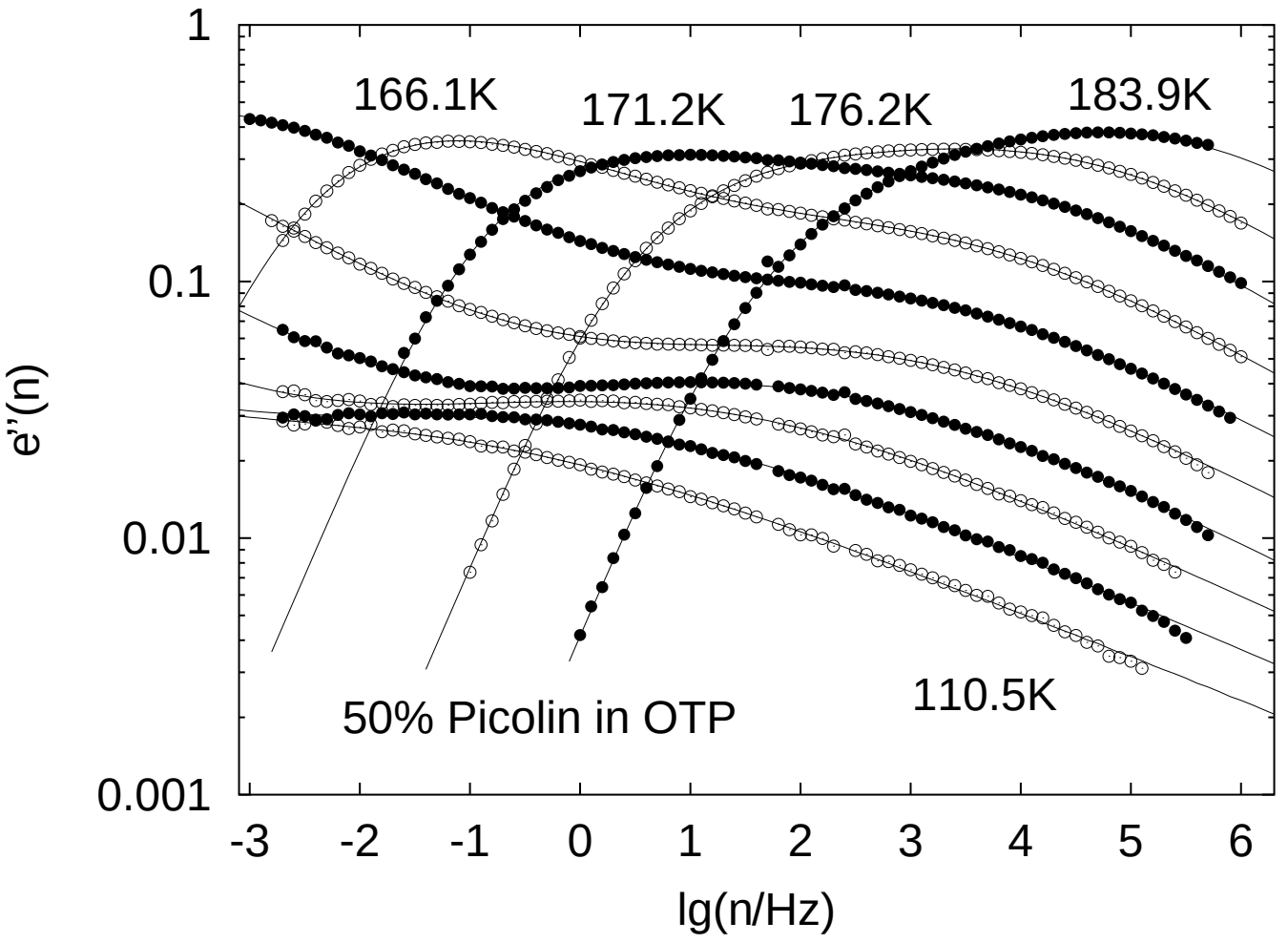

**Figure 12:** Dielectric loss spectra of the binary glass former 50 weight-% 2-picoline in o-terphenyl (OTP); interpolation applying the distributions $G_{GG}$ and $G_\beta$ in the frame of the Williams-Watts ansatz (solid lines).